# An Extended Galactic Population of Low-Luminosity X-Ray Sources (CVs?) And The Diffuse X-Ray Background

Eyal Maoz and Jonathan E. Grindlay

Harvard-Smithsonian Center for Astrophysics,
MS 51, 60 Garden Street, Cambridge, MA 02138

E-mail: maoz@cfa.harvard.edu ; josh@cfa255.harvard.edu

## ABSTRACT

The incompatibility of the properties of the x-ray background (XRB) with AGNs contributing $\gtrsim 60\%$ at energies of a few KeV has often been interpreted as being due to a substantial contribution of a new population of yet unrecognized x-ray sources. The existence of such population has been recently suggested also by an analysis of very deep ROSAT observations (Hasinger *et al.* 1993) which revealed a considerable excess of faint x-ray sources over that expected from QSO evolution models, and that the average spectrum of the resolved sources becomes harder with decreasing flux limit. These sources could be extragalactic in origin, but if they make a substantial contribution to the XRB then they must exhibit much weaker clustering than galaxies or QSOs in order to be consistent with the stringent constraints on source clustering imposed by autocorrelation analyses of the unresolved XRB.

We investigate the possibility that the indicated new population of x-ray sources is Galactic in origin. Examining spherical halo and thick disk distributions, we derive the allowed properties of such population which would resolve the discrepancy found in the number counts of faint sources, and be consistent with observational constraints on the total background intensity, the XRB anisotropy, the number of unidentified bright sources, the Galaxy's total x-ray luminosity, and with the results of fluctuation analyses of the unresolved XRB.

We find that a flattened Galactic halo (or a thick disk) distribution with a scale height of a few Kpc is consistent with all the above requirements. The typical x-ray luminosity of the sources is $\approx 10^{30-31}$ erg/s in the 0.5-2 KeV band, the number density of sources in the Solar vicinity is $\sim 10^{-4.5}$ pc$^{-3}$, their total number in the Galaxy is $\sim 10^{8.5}$, and their total contribution to the Galaxy's x-ray luminosity is $\sim 10^{39}$ erg/s.

We discuss the possible nature of these sources, including them being subdwarfs, LMXBs, and old neutron stars. We argue that the inferred x-ray and optical luminosities of the sources, their $\sim$ 2-4 KeV spectrum, and the derived local number density and spatial distribution are all consistent with them being intrinsicly faint cataclysmic variables with low accretion rates. We suggest a few possibilities for the origin of such population, including an origin from disrupted globular clusters or dark clusters. We make predictions and suggest tests that could either confirm or rule out our proposal in the near future.

## 1. INTRODUCTION

More than 30 years after the discovery of the x-ray background (Giacconi *et al.* 1962) the origin of this radiation is still not entirely understood. Mather *et al.* (1990) have shown that thermal Bremsstrahlyung from a diffuse hot medium cannot provide a significant fraction of the XRB, thus leaving the alternative possibility that the XRB arises predominantly from an unresolved population of discrete sources. This is also supported by fluctuation analyses which show that the contribution of discrete sources to the $\simeq$1-2 KeV background is $\gtrsim$ 75-90%, and it is consistent with the entire XRB being due to discrete sources (*e.g.*, Hamilton & Helfand 1987; Soltan 1991; Shanks *et al.* 1991; Hasinger *et al.* 1993).

An important step towards resolving the XRB has been recently made by Hasinger *et al.* (1993) who analyzed a 152 Ksec ROSAT PSPC observation in the direction of the absolutely lowest $H_I$ column density (the Lockman Hole), as well as 26 shallower fields from the ROSAT Medium Sensitivity Survey. Hasinger *et al.* (1993) found that $\sim$60% of the 1-2 KeV background is resolved into discrete sources with flux $\geq 2.5\times10^{-15}$ cgs in the 0.5-2 KeV band (hereafter, cgs stands for erg cm$^{-2}$ s$^{-1}$). The logarithmic slope of the differential source counts is $\sim 2.7 \pm 0.3$ at bright fluxes and $\sim 1.9 \pm 0.2$ at faint fluxes, which confirms the flattening of the logN-logS relation at faint fluxes that has been suggested by earlier studies (Hasinger *et al.* 1991; Shanks *et al.* 1991).

While the shape of the logN-logS relation is consistent with that expected from evolutionary models for the AGN x-ray luminosity function (XLF), Hasinger *et al.* (1993) draw the attention to a significant difference in the number counts at the faintest flux limit. The number of observed sources with flux $\geq 2.5\times10^{-15}$ cgs is 1.6 times larger than the predicted by XLF models (Boyle *et al.* 1993). Hasinger *et al.* (1993) suggest that this discrepancy may be resolved by adopting a more complicated model for the QSO XLF evolution, or alternatively, by the existence of a new population of sources with a steep logN-logS relation at low fluxes. The later possibility is especially interesting in light of the finding that the average spectrum of the resolved sources becomes harder with decreasing flux. Hasinger *et al.* (1993) find that at fluxes below $10^{-14}$ cgs there is a substantial fraction of objects with harder spectrum ($3.3\sigma$), which can be interpreted as an additional indication for the increasing fraction of a new population of sources at low fluxes.

Indeed, it has been previously suggested that a yet unrecognized class of x-ray sources may make a non-negligible contribution to the XRB. These include intrinsicly absorbed AGNs (Grindlay & Luke 1990; Hasinger *et al.* 1993), low-luminosity AGNs (Setti 1992), and star-forming galaxies at moderate redshifts (Griffiths & Padovani 1990). However, there is some difficulty in having the entire XRB be due to extragalactic populations. Autocorrelation analyses fail to find any clear signal for a cosmic origin in the unresolved background (Barcons & Fabian 1988, 1989; Soltan 1991; Hasinger *et al.* 1991; Wang & McCray 1993). Fluctuation analyses (Barcons & Fabian 1988; Persic *et al.* 1989; Carrera & Barcons 1992; Carrera *et al.* 1993; Soltan & Hasinger 1994) *severely* constrain the clustering properties of sources producing the XRB, regardless of their nature. In particular, they have shown that sources which are clustered like galaxies or like the optically observed QSOs cannot provide more than about 60-65% of the total background intensity. This supports earlier claims by Hamilton and Helfand (1987) that no reasonable extrapolation of the known population of x-ray emitting quasars can compose the entire XRB, and that a new population of numerous, low-luminosity x-ray sources is required.

The constraint on the clustering properties of the sources making the XRB may eliminate some of the conceivable candidates for the faint sources, but does not rule out an extragalactic origin altogether (*e.g.*, faint blue galaxies are weakly clustered [Efstathiou *et al.* 1991]). In the present paper we investigate the interesting possibility that the discrepancy in the faint x-ray source counts (Hasinger *et al.* 1993) is due to an as yet unrecognized *Galactic* population which makes a non-negligible contribution to the soft XRB, and which would naturally be consistent with the clustering constraint. We derive the characteristics of the proposed Galactic population that are allowed by various observational constraints (§§2,3), discuss the possible nature and origin of these sources, and argue that they are most likely to be intrinsicly faint cataclysmic variables (§4). In §5 we summarize the results, make predictions, and propose tests.

## 2. CONSTRAINTS ON A GALACTIC POPULATION

We first investigate whether a Galactic population of x-ray sources that has some "reasonable" spatial distribution could resolve the discrepancy found in the number counts of faint sources (§1), while being consistent with several observational constraints: its contribution to the residual XRB; its contribution to the number of unidentified bright sources; its integrated contribution to the Galaxy's x-ray luminosity, and limits from fluctuation analyses on the surface density of sources which make the unresolved background. Assuming that the sources are standard candles with luminosity $L$ at the relevant energy band, we shall examine two models for the source distribution: spherical halo models with a varying core radius and an asymptotic $r^{-2}$ density profile, and exponential thick disk models with varying scale height. At this stage we shall not consider other conceivable spatial distributions and luminosity functions because we wish to constrain the *typical* properties of the proposed population, and not allow the flexibility of having too many degrees of freedom.

### 2.1. Spherical Halo Distributions

Let us assume a population of x-ray sources with a number density $n(R)$ given by

$$n(R) = \frac{n_0(R_c^2 + R_0^2)}{(R_c^2 + R^2)} \quad , \tag{1}$$

where $R_c$ is the core radius of the distribution, $R_0 = 8.5$ Kpc is the Solar distance to the Galactic center, R is a Galactocentric distance, and $n_0$ is the number density of the sources in the Solar vicinity. The surface number density of sources with flux greater than $S$ in a direction which makes an angle $\theta$ with the direction to the Galactic center is

$$\frac{dN(>S,\theta)}{d\Omega} = \int_0^{(L/4\pi S)^{1/2}} n(r,\theta)\, r^2 dr \quad , \tag{2}$$

where $r$ is the distance from the Sun in that direction, and

$$n(r,\theta) = \frac{n_0(1 + a^2)}{1 + a^2 + x^2 - 2ax\cos\theta} \quad , \tag{3}$$

where $x \equiv r/R_c$, and $a \equiv R_0/R_c$. Substituting Eq. (3) in Eq. (2) we obtain

$$N(>S_{14},\theta) = \frac{0.187(1+a^2)n_0}{a^3}\left\{x + a\cos\theta\log\left[\frac{1+a^2+x^2-2ax\cos\theta}{1+a^2}\right] + \right.$$

$$\left.\frac{2a^2\cos^2\theta-(1+a^2)}{(1+a^2\sin^2\theta)^{1/2}}\left\{\arctan\left[\frac{x-a\cos\theta}{(1+a^2\sin^2\theta)^{1/2}}\right] - \arctan\left[\frac{-a\cos\theta}{(1+a^2\sin^2\theta)^{1/2}}\right]\right\}\right\} \quad \mathrm{deg}^{-2} \quad , \tag{4}$$

where $x \equiv 0.38a(L_{31}/S_{14})^{1/2}$, $L_{31}$ is the source luminosity in units of $10^{31}$ erg/s, and $S_{14}$ is the observed flux in units of $10^{-14}$ cgs. Hasinger *et al.* (1993) find the number density of sources with flux $\geq 2.5\times10^{-15}$ cgs in the 0.5-2 KeV band to be $\simeq 413$ deg$^{-2}$, and point out that it exceeds by $\simeq 60\%$ the expected number from QSO evolution models (Boyle *et al.* 1993). They suggest that this discrepancy may be resolved either by adopting a more complicated evolution model for the XLF, or by the existence of a new population of sources, which is the possibility addressed in the present paper.

Optical observations (Hasinger *et al.* 1993) and spectroscopic identifications (Shanks *et al.* 1991) indicate that stars contribute $\lesssim 10\%$ to the number counts at this flux level. Thus, assuming that their fraction is 10%, the above numbers imply a surface density of $\approx 120$ deg$^{-2}$ non-QSO, non-stellar sources at this flux limit. Requiring that the proposed population produce this number of counts provides the first constraint on the free parameters ($L$,$n_0$,$R_c$).

The number counts of Galactic sources depend on direction, so equation (4) should be evaluated in the direction where the observations were performed. The results of Hasinger *et al.* (1993) are based on an analysis of 27 fields at high Galactic latitude ($|b| > 30°$) and which are at angular distances of $60° \lesssim \theta_i \lesssim 120°$ from the Galactic center ($i = 1,..,27$). Therefore, we shall average equation (4) over those fields, and require that $N(>0.25) = 120$, where

$$N(>S_{14}) = \frac{1}{27}\sum_{i=1}^{27} N(>S_{14},\theta_i) \quad . \tag{5}$$

This requirement provides a relation between $L$ and $n_0$ for a given spatial distribution, and this relation is presented in by the solid line in Figure 1 for two very different halo models. The various fields should, in principle, have had unequal weights in the above averaging since they differ considerably in exposure time and in the $H_I$ column density along the line of sight (Hasinger *et al.* 1993). Still, equation (5) is a reasonable first approximation for the effective $N(>S)$ since the distribution of $\theta_i$ is within a limited range, and it is peaked around $\simeq 90°$.

The fraction of unidentified sources in x-ray samples decreases with increasing flux limit. Thus, the fact that the proposed class of sources is so far unrecognized places a limit on its contribution to the source counts at higher flux levels where optical identifications are much more complete. An estimate for the density of unidentified sources with flux $\geq 10^{-14}$ cgs in the 0.5-2 KeV band is provided by several analyses (Boyle *et al.* 1993; Shanks *et al.* 1991; Hasinger *et al.* 1991) of Deep ROSAT PSPC observations of the fields QSF1 and QSF3 (Boyle *et al.* 1990). Both fields are at high galactic latitude and at an angular distance $\theta \simeq 102°$ from the Galactic center. These investigations find a density of $\sim 90$ deg$^{-2}$ non-stellar sources of which $63 \pm 15$ deg$^{-2}$ are identified QSOs. Since the fraction of stars at this flux limit is estimated to be $\sim$ 10-15% (Hasinger *et al.* 1991; Shanks *et al.* 1991), the above numbers imply a total source density of $\approx 108$ deg$^{-2}$, of which $\simeq 25\%$ are unidentified (about half of those do not even have an optical counterpart with $B < 22.3$ mag [Shanks *et al.* 1991]). Hasinger *et al.* (1993) find a similar total number density of $\simeq 110\,\mathrm{deg}^{-2}$ sources with flux $\geq 10^{-14}$ cgs (0.5-2 KeV).

Therefore, a constraint on the characteristics of the proposed population is that its contribution to the number density of unidentified sources with fluxes $\geq 10^{-14}$ cgs would not exceed 25% of the total number of sources observed at this flux limit, i.e., we require that $N(> 10^{-14})$, as defined in equation (5), would not exceed 25% of $110\,\mathrm{deg}^{-2}$. The shaded area in Figure 1 represents the combinations of $L$ and $n_0$ which yield a contribution between 15-25% to the number counts at the above flux limit. One should bear in mind that this is not a very well defined constraint because it is unclear how well the population of unidentified sources corresponds to the proposed class of sources. Source confusion may lead to some false identifications, especially if the proposed sources are very faint in the optical band and have a typical $F_x/F_{opt}$ ratio which is much different than that of known classes of sources.

So far we discussed constraints on number counts of resolved sources, but there is also an observational constraint on the contribution of the *unresolved* fraction of the proposed population to the unresolved XRB. The contribution of the entire proposed population to the XRB intensity, averaged over the directions to the fields studied by Hasinger *et al.* (1993) is given by

$$\left\langle \frac{dI}{d\Omega} \right\rangle = \sum_{i=1}^{27} \int_0^{r_{max}(\theta_i)} n(r,\theta_i)\, \frac{L}{4\pi r^2}\, r^2\, dr \quad , \tag{6}$$

where $r_{max}(\theta_i)$ is derived by assuming $R_{max} = 100$ Kpc. Hasinger *et al.* (1993) found that about 60% of the background is resolved to discrete sources with total flux that amounts to $1.48 \times 10^{-8}$ cgs $\mathrm{sr}^{-1}$ in the 0.5-2 KeV band. This implies a total background intensity, averaged over the analyzed fields, of $2.47 \times 10^{-8}$ cgs $\mathrm{sr}^{-1}$. The contribution of the proposed population to the unresolved XRB, relative to the contribution of AGNs, depends on the unknown relative behavior of their logN-logS relations at fluxes below the current detection limit. However, since $\sim 30\%$ of the sources at the faintest flux limit are proposed to belong to the new population, it is reasonable to expect their entire contribution to the XRB intensity in this band to be $\approx$20-40%. Thus, requiring that equation (6) yield 20-40% of the background intensity constrains the parameters $L$ and $n_0$ to the region between the two dashed lines in Figure 1.

Fluctuation analyses of the unresolved background reveal that the entire background can be produced by a population of discrete sources, and provide a constraint on the number density of unresolved sources. Hamilton and Helfand (1987) obtained a total number density of $\sim 5000\,\mathrm{deg}^{-2}$ (for fields at $|b| > 30°$), Shanks *et al.* (1991) obtained a sky density of $\sim 4000\,\mathrm{deg}^{-2}$, and Soltan (1991) concluded a density of $\sim$1500-4800 $\mathrm{deg}^{-2}$. Thus, requiring that the total number density of the proposed population, averaged over the fraction of the sky with $|b| > 30°$, be between 1000-4000 $\mathrm{deg}^{-2}$, constrains the values of $L$ and $n_0$ to the region between the two horizontal dash-dot lines in Figure 1.

The total contribution of the proposed population to the x-ray luminosity of the Galaxy is given by

$$L_{gal} = L \int_0^{R_{max}} 4\pi n(R)\, R^2 dR \;=\; \frac{4\pi n_0 L R_0^3}{a^3} \left[ \frac{a\, R_{max}}{R_0} - \arctan\left( \frac{a\, R_{max}}{R_0} \right) \right] \quad . \tag{7}$$

Although the total Galaxy luminosity is unknown, we know that other spiral galaxies have total luminosities between $\sim 10^{38}$ to a few times $10^{41}$ erg/s in the (0.2-3.5) KeV band (Fabbiano 1989). Thus, a reasonable upper limit for the luminosity of the proposed population in the 0.5-2 KeV band would be $\sim 10^{40}$ erg/s. Substituting $R_{max} = 100$ Kpc in equation (6), the region of the $(L, n_0)$ parameter space which is below the dotted line in Figure 1 corresponds to models satisfying $L_{tot} \leq 10^{40}$ erg/s.

We shall now repeat the above analysis for thick disk distributions, and then discuss the implications of the results for both types of models. The issue of anisotropy in the x-ray emission due to the proposed Galactic population is discussed in detail in §3.

### 2.2. Thick Disk Distributions

The space densities of various types of stars perpendicular to the galactic plane can be fit by an exponential with a scale that varies from 50 pc for O stars to 2 Kpc for Subdwarfs (Mihalas & Binney 1981). The radial profile of these distributions can be fit by an exponential with scale $R_d \sim 5 \pm 1$ Kpc (Gilmore, King & van der Kruit 1990). Thus, we shall assume that the proposed population of Galactic x-ray sources can also be reasonably well fit by

$$n(R, z) = n_0 \exp\left(-\frac{|z|}{R_z} - \frac{(R - R_0)}{R_d}\right) \quad . \tag{8}$$

where

$$R = \left(R_0^2 + r^2 \cos^2 b - 2R_0 r \cos b \cos l\right)^{1/2} \quad , \tag{9}$$

$R$ and $z$ are the conventional cylindrical coordinates, $r$ is the distance from the sun, and $z = r \sin b$. For simplicity we assume $R_d = 5$ Kpc, thus leaving the scale height in the $\hat{z}$ direction, $R_z$, and the number density of the sources in the Solar vicinity, $n_0$, as the free parameters. Clearly, $R_d$ need not necessarily be identical to that of the observed stars, and the scale height may vary with $R$. But, for our purpose, which is to constrain the *typical* characteristics of the proposed population, it may provide a reasonable first approximation for a thick disk distribution.

We have exactly repeated the analysis described in §2.1, except that now we replaced $\theta_i$ by $(l_i, b_i)$. The various constraints on $L$ and $n_0$ for thick disk distributions are presented in Figures 1-c, and 1-d.

### 2.3. Allowed Characteristics Of The Proposed Population

Figure 1 presents all the constraints (see §2.1) on the typical x-ray luminosity of the sources in the 0.5-2 KeV band , $L$, and on their number density in the Solar vicinity, $n_0$, for two spherical halo models with small and large core radii, and for two thick disk models with modest and large scale heights. For each model, the solid line represents the combinations of $L$ and $n_0$ that would exactly resolve the discrepancy found in the faint source counts; the shaded area corresponds to combinations which would produce a number density of unidentified bright sources which is consistent with observations; the two inclined dashed lines confine the parameter space which would not violate constraints on the contribution to the XRB intensity; the horizontal dash-dot lines confine the region which would be consistent with the results of fluctuation analyses of the unresolved background, and the region below the dotted line corresponds to models which contribute $\leq 10^{40}$ erg/s to the Galaxy's total x-ray luminosity.

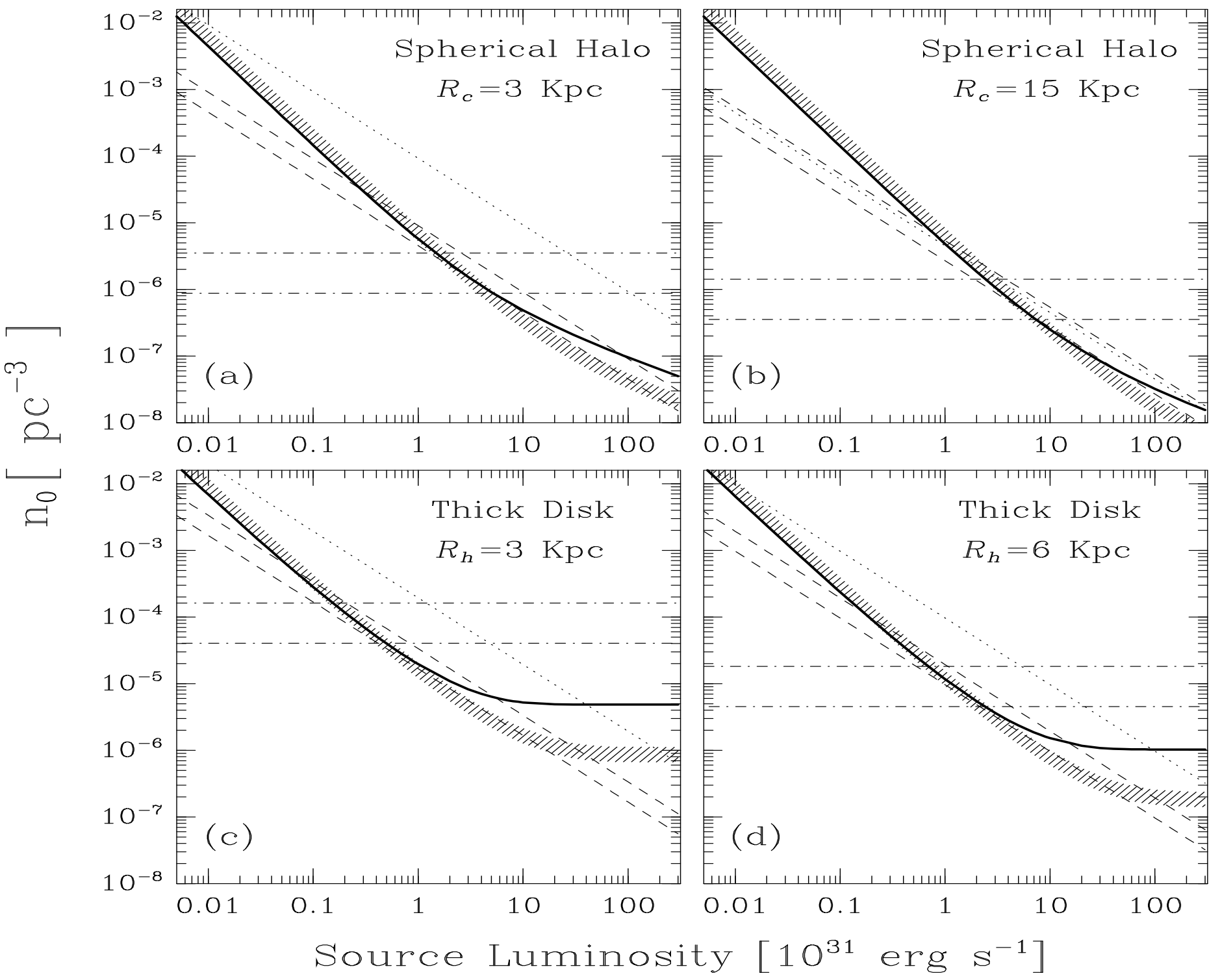

Fig. 1.— Combined observational constraints on the typical x-ray luminosity of the sources, $L$, and their number density in the Solar vicinity, $n_0$, for two different halo models (a,b) and thick disk models (c,d) (see detailed discussion in §2.3).

It is interesting to notice that all the constraints are satisfied within a well defined region in parameter space, a situation which need not necessarily happen. The typical luminosity of the sources is $10^{31\pm0.5}$ erg/s (0.5-2 KeV), and it is not very sensitive to changes in the scale of the spatial distribution. Distributions of sources with much higher or much lower typical x-ray luminosities cannot be consistent with all the constraints and still resolve the discrepancy in the faint source counts. Table 1 presents the typical characteristics of a few models which are consistent with all the constraints discussed in §2.1, and Figure 2 presents the logarithmic slope of the differential logN-logS relation for these models which was calculated using equation (5). All models have an approximately Euclidean logN-logS relation at fluxes above $\sim 10^{-14}$ cgs , and which flattens considerably below a flux of a few times $10^{-16}$ cgs . In the flux range of $10^{-15}$-$10^{-14}$ cgs the logN-logS relation is steeper than the expected for an evolving population of QSOs (Boyle *et al.* 1993), thus providing an increasing relative contribution of the Galactic sources with decreasing flux.

| | Thick Disk ($R_d = 5\,$Kpc) | | Spherical Halo | |
|---|---|---|---|---|
| | $R_h = 3\,$Kpc | $R_h = 6\,$Kpc | $R_c = 3\,$Kpc | $R_c = 15\,$Kpc |
| $L$ [erg/s] | $2.8\times10^{30}$ | $7.0\times10^{30}$ | $1.5\times10^{31}$ | $2.4\times10^{31}$ |
| $n_0$ [pc$^{-3}$] | $8\times10^{-5}$ | $1.8\times10^{-5}$ | $3\times10^{-6}$ | $1.4\times10^{-6}$ |
| $L_{total}$ | $1.1\times10^{39}$ | $1.4\times10^{39}$ | $4.5\times10^{39}$ | $7.2\times10^{39}$ |
| $N_{total}$ | $4\times10^{8}$ | $2\times10^{8}$ | $3\times10^{8}$ | $3\times10^{8}$ |
| $N(S \geq 10^{-11}$cgs) | 38 | 34 | 18 | 17 |

**Table 1** - The typical x-ray luminosity of the sources, $L$, and the number density of sources in the Solar vicinity, $n_0$, for two thick disk distributions with exponential disk scale of 5 Kpc and scale height $R_h$, and two spherical halo distributions with core radius $R_c$ (see Figure 1 and §2.3). Also shown are the total number of sources in the Galaxy, $N_{total}$, their integrated luminosity, $L_{total}$, and the expected number of nearby sources with flux above $10^{-11}$ cgs (see §4.1). All x-ray luminosities correspond to the 0.5-2 KeV band.

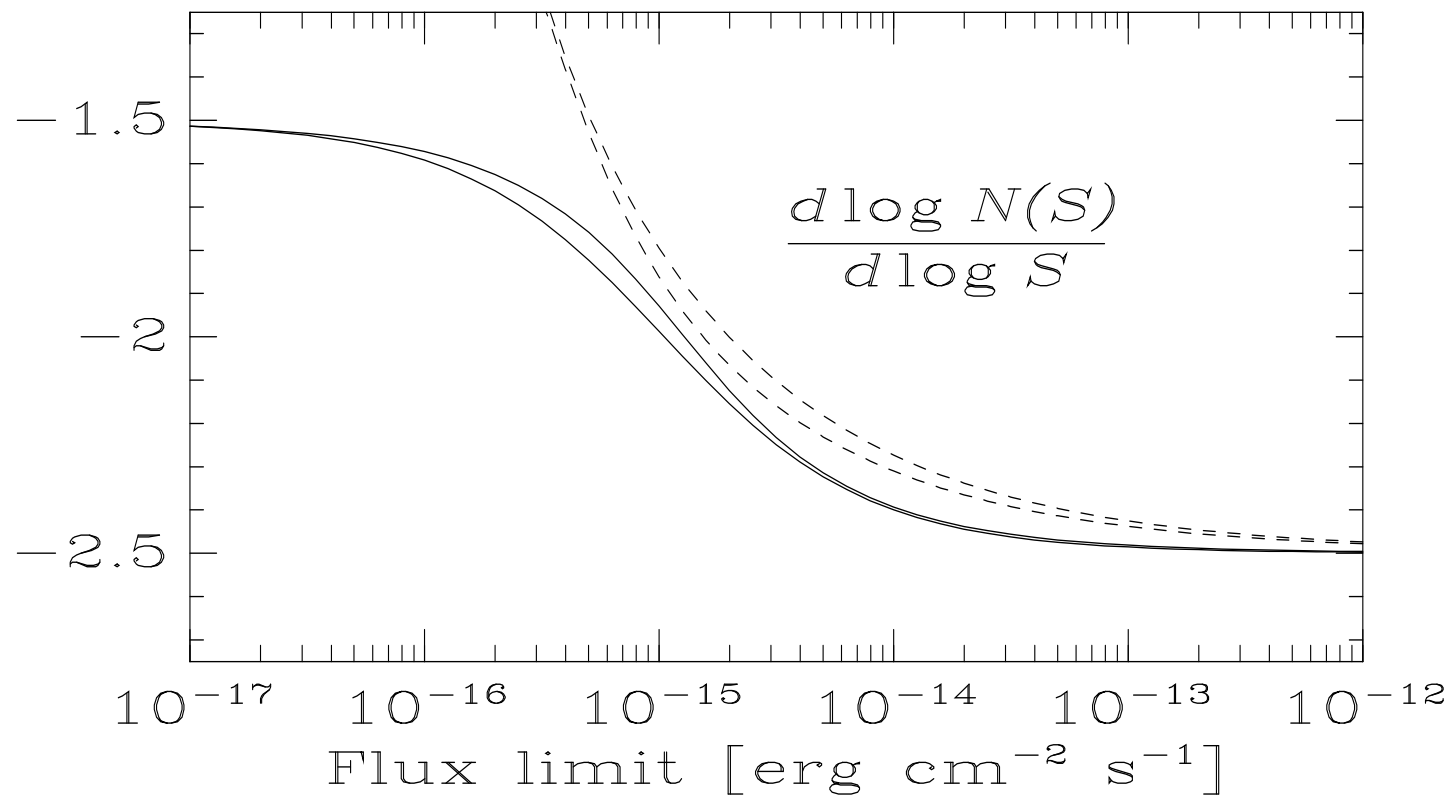

Fig. 2.— The logarithmic slope of the differential logN-logS relation for the two spherical halo models (solid curves), and the two thick disk models

## 3. SPECTRUM AND ANISOTROPY

We now examine whether the existence of the proposed population is consistent with the observed anisotropy of the XRB. Since the anisotropy is energy-band dependent, we first need to have some information about the spectrum of the sources.

The spectrum of the XRB in the 3-10 Kev range can be fit by a power law with an energy spectral index of $\alpha$=0.4 (*e.g.*, Boldt 1987; Wu *et al.* 1991; Carrera & Barcons 1992). Below that energy the spectrum lies well above the extrapolation of Boldt's formula, and a simple power-law fit gives $\alpha \sim 0.7$ in the $\sim$ 0.5-3 KeV range. Apparently, the background intensity exceeds the extrapolation of the $>$ 3 KeV spectrum by $\sim 30\%$ at 1-3 KeV (Wu *et al.* 1991), and by a factor of 2 at 0.5-1 KeV (Shanks *et al.* 1991). This may simply reflect the shape of the spectrum of extragalactic sources, but it could alternatively be due to some population of sources, either galactic or extragalactic (McCammon & Sanders 1990) which contribute to the XRB especially at these energies.

The resolved sources with flux above $10^{-14}$ cgs (0.5-2 KeV), of which most belong to known types of extragalactic sources, have an average spectral index of $\alpha \simeq 1.3 \pm 0.2$ in this band (Shanks *et al.* 1991; Hasinger *et al.* 1991). Since it is steeper than the background spectrum the sources with flux $<10^{-14}$ cgs , combined with the unresolved part of the XRB must have a harder average spectrum. This means that if the XRB is mainly due to discrete sources then the faint sources must have harder spectrum than the XRB (Boldt & Leiter 1984). Carrera and Barcons (1992) subtracted the extrapolated contribution of the detected source counts and the expected contribution of clusters of galaxies from the background, and obtained a rough estimate of $\alpha \sim 0.3$ for the spectral index of the remaining XRB in the 0.6-2.4 KeV band. Indeed, it has been recently discovered (Hasinger *et al.* 1993) that the average spectrum of the resolved sources becomes harder with decreasing flux; the average energy spectral index of *all* resolved sources with flux $\geq 2.5\times10^{-15}$ cgs is $\alpha = 0.96\pm0.11$ in the 0.5-2 KeV band, which may imply an average spectral index of $\sim$0.3-1.0 for the proposed population of sources alone. Assuming a thermal Bremsstrahlyung emission, this implies $KT \sim 2$-4 KeV.

We expect the proposed population of sources to introduce an energy-dependent degree of anisotropy to the XRB due to our offset position with respect to the Galactic center, as well as due to the possible flattened shape of the spatial distribution of the sources (§2). Indeed, there is already evidence for Galactic components at various energy bands. In the 2-60 KeV band, and in particular in the 2-10 KeV band, two Galactic emission components have been clearly detected. One is a thin disk component which is a few degrees wide and has a radial extent of $\sim 10$ Kpc (Bleach *et al.* 1972; Worrall *et al.* 1982; Kahn & Caillault 1986). The second component has much larger scale and extends to high Galactic latitudes. It is best modeled by a finite radius thick disk with a scale height of $R_z \approx 3$-6 Kpc, and a radial scale of $R_d \approx 12$-20 Kpc (Iwan *et al.* 1982). A spherical halo model is less successful in fitting the data. Iwan *et al.* (1982) emphasis that the effects of the local supercluster, the North Polar Spur, and the Compton-Getting effect cannot account for this anisotropy. This component contributes $\sim 10\%$ to the 2-60 KeV background in the direction of the Galactic center. The source of this emission is unclear, but it can be due a population of discrete sources (McCamon & Sanders 1990).

The fraction of Galactic contribution to the XRB increases with decreasing energy: the background intensity in the 0.16-3.5 KeV band strongly depends on Galactic longitude even when only high latitude data are used (Wu *et al.* 1991). This has been interpreted as a clear indication for a large-scale Galactic component which is not confined to local features such as the North Polar Spur (Wu *et al.* 1991). In the 0.5-1 KeV range the anisotropy is dominated by an irregular feature about 110° in diameter located in the general direction of the Galactic center and which coincides with the general structure of the radio Loop I.

Evidently, there is at least one Galactic component to the XRB with an unclear origin, and it is likely to have a flattened spatial distribution, e.g., a thick disk or an oblate spheroid. Thus, the global structure of the background anisotropy which is expected due to the proposed Galactic population is not in conflict with observations. However, it is difficult at this stage to determine whether the amplitude of the large-scale variations is consistent with the proposed population for two reasons. First, it requires knowing the average contribution of the Galactic sources to the XRB in the specific energy bands in which the degree of large-scale anisotropy has been measured. Although we have estimated a source temperature of 2-4 KeV, these sources may not have a simple thermal spectrum (*e.g.*, cataclysmic variables may have both soft and hard components). Second, the energy bands in which the anisotropy has been detected are quite broad, and this tends to smear out the anisotropic signal: most of the XRB energy comes from high energy photons, so an anisotropic component with typical energies of a few KeV will almost disappear in the 2-60 KeV band. Similar problem would occur in energy bands which go much below $\sim$1 KeV (*e.g.*, the 0.16-3.5 KeV range discussed above). At low energies the Galaxy becomes considerably opaque due to photoelectric absorption (McCamon & Sanders 1990), thus any anisotropic contribution in these energies due to an extended Galactic distribution will look almost isotropic. This is especially true if the Galactic sources have a flattened distribution, as does the $H_I$, so that the column density of $H_I$ in every direction is proportional, to first order, to the emission of the proposed population in that direction.

## 4. NATURE OF THE SOURCES

Most known classes of galactic x-ray sources with large scale heights seem to have space densities much lower than the required (Table 1). For example, the distribution of Subdwarfs (*e.g.*, dM stars) has a scale height of $\sim 2$ Kpc, but a local space density of $\sim 1.6\times10^{-6} M_\odot$ pc$^{-3}$ (*e.g.*, Allen 1973) which is $\sim 1\%$ of the required value (Figure 1-c). In addition, their x-ray luminosity is $\lesssim 10^{29}$ erg/s (Johnson 1981; Schmitt *et al.* 1990), *i.e.*, below the required luminosity of $10^{30-31}$ erg/s, and their spectra are typically softer than the inferred for our sources.

Low-mass x-ray binaries (LMXBs) have typical x-ray luminosities of $L_x \sim 10^{35.5-37.5}$ erg/s, but quiescent LMXBs have $L_x \sim 10^{32-34}$ erg/s (Verbunt *et al.* 1994; Grindlay 1994a). It is not inconceivable that LMXBs with very low accretion rates could have $L_x \sim 10^{31}$ erg/s which would make them appealing candidates for the proposed population, especially if they have $L_x/L_{opt}$ ratio similar to the 10-$10^4$ ratios for active LMXBs (Bradt & McClintock 1983). However, while active LMXBs have hard spectrum (Verbunt 1993) that could be consistent with the required $\lesssim 5$ KeV spectrum for the proposed population (§3), LMXBs with very low accretion rates typically have a much too soft spectrum which corresponds to KT$\sim$0.1-0.3 KeV (Verbunt *et al.* 1994). This property of LMXBs with low accretion rate can also be understood theoretically (Ramesh Narayan, private communication). Thus, it seems that accreting neutron stars in binary systems can be excluded as candidates for the proposed population.

We shall now argue that a population of low-luminosity cataclysmic variables is consistent with all the required properties, and that it is also indicated by independent observations. Before discussing the possible origin of these CVs (§4.2) we shall briefly discuss an exciting but much more speculative possibility for the nature of these sources (§4.3).

### 4.1. Intrinsicly Faint Cataclysmic Variables

Cataclysmic variables (hereafter CVs) are x-ray sources with a few KeV spectrum (although a single temperature Bremsstrahlyung emission may not always fit well their spectra at energies of a few KeV). Patterson and Raymond (1985) studied the correlations between the x-ray luminosity of CVs, their x-ray-to-visual flux ratio, and their accretion rate. Their results indicate that CVs with very low accretion rates ($\lesssim 10^{15}$ g/sec, or $1.6\times10^{-11}M_{\odot}$ yr$^{-1}$) are strong x-ray emitters with $L_x\sim10^{30}$-$10^{31}$ erg/s in the 0.2-4.0 KeV band, faint in the optical band ($M_v\sim$11-12), and have $L_x/L_v\sim$1-10. The $L_x/L_v$ ratio goes *down* with increasing accretion rate, a trend which is attributed to the fact the high accretion rates imply an optically thick disk where a substantial fraction of the x-ray luminosity is reprocessed. Very low accretion rate systems have a small optically thin disk and consequently low optical luminosity. Patterson and Raymond (1985) show that CVs with accretion rate as low as $10^{-12}M_{\odot}$ yr$^{-1}$ have $M_v\sim$13-14 and $L_x\sim10^{30}$. Thus, the spectrum and x-ray luminosity of low-$\dot{\rm M}$ CVs are consistent with our results for the proposed population (§2.3), and they are faint enough to appear totally uninteresting in the optical band, except for the very nearby ones. But, does the observed local number density of such CVs coincide with our predictions (Table 1)?

Surprisingly, this fits nicely as well. Hertz *et al.* (1990) report on four spectroscopically identified CVs, originally discovered and thus selected by an x-ray survey (Hertz & Grindlay 1984), with $L_x\lesssim10^{31}$ erg/s, and which are optically fainter then other known CVs ($M_v>11.5$). Two of them are nearby ($d<200$ pc), and have a very low inferred mass accretion rate ($\dot{M}<2\times10^{-12}M_{\odot}$ yr$^{-1}$). Hertz *et al.* (1990) argue that these CVs belong to a large population of low-luminosity CVs that release the majority of their accretion energy as x-rays ($F_x/F_{opt}>5$). They emphasize that it would be extremely difficult to discover these objects through traditional means (except through their x-ray emission or very blue colors). They derive a local space density of $\gtrsim$2-3$\times10^{-5}$ pc$^{-3}$ for these CVs, a density which is 3-5 times higher than densities estimated from optical surveys of CVs (Patterson 1984; Downes 1986). An optical survey (Shara, Moffat, & Potter 1990) revealed blue CVs in numbers that imply a space density as high as $10^{-4}$ pc$^{-3}$. Thus, independent observations indicate a local number density of $10^{-5}$-$10^{-4}$ pc$^{-3}$ for low-$\dot{\rm M}$ CVs, which is consistent with our results (Table 1, and §2.3).

The observed population of intrinsically faint CVs is not confined to the Galactic plane. Howell and Szkody (1990) studied a sample of CVs at high galactic latitude and found that most of them are extremely faint. Assuming typical CV luminosities yield distances of 0.35-8 Kpc, thus suggesting that these are members of the Galactic halo and may represent a distinct population of CVs. However, Howel and Szkody (1990) also point out an alternative possibility that these are nearby members of a distribution of intrinsically very faint systems ($M_v>10$), which would also imply that these objects have a considerable space density. This idea coincides with the conclusion of Hertz *et al.* (1990), and the inferred optical luminosity is indeed consistent with a low-$\dot{M}$ CV (see discussion above).

Howell and Szkody (1990) also find that the faint CVs show preference for orbital periods below the period gap, and point out that the significantly lower mean orbital period implies that these systems are relatively old. If old CVs are detectable in the x-ray at the current epoch then they are likely to have a low accretion rate, and thus a low x-ray luminosity (Peterson & Raymond 1985) which coincides with our prediction (Table 1). It is interesting to notice that magnetic CVs (*e.g.*, AM-Her type) are also predominantly below the period gap, and are also optically fainter than nonmagnetic CVs (larger $L_x/L_v$ ratio) since they do not have well developed accretion disks. Thus, it is possible that the proposed population of x-ray sources are some kind of magnetic CVs.

Before we discuss the possible origin of these systems let us perform one additional consistency check. Assuming that we are indeed immersed within an extended Galactic distribution of low-luminosity CVs, we should be able to detect a few relatively nearby ones at relatively high fluxes. The number of objects which belong to the proposed population and should be nearby enough to be observed with flux $\gtrsim 10^{-11}$ cgs is around 17-38 within $4\pi$ steradians (Table 1). Indeed, Optical identifications of the Einstein Observatory EMSS data (Gioia *et al.* 1990), which covers 778 deg$^2$ (1.9% of the sky area), include two CVs with flux $\gtrsim 10^{-11}$ cgs . Although it is a small number statistics, there is no apparent shortage of CVs at this flux limit.

### 4.2. Origin Of An Extended CV Distribution

If the proposed picture is confirmed by subsequent investigations (see predictions in §5) then understanding the origin of these CVs may provide valuable information on various aspects of Galaxy formation. It is very unlikely that any Galactic disk population could be heated by gravitational interactions to heights of a few Kpc above the plane to form a thick disk distribution (*e.g.*, Jenkins & Binney 1990). It is more plausible that the proposed CV population formed during the dissipational collapse of the Galaxy, and thus ended up in a thick disk distribution in a similar way to the formation of other stellar populations with large scale heights (*e.g.*, Gilmore 1984; Sandage & Fouts 1987). It was also shown that a large fraction of the halo dark matter could be in the form of white dwarfs (Ryu, Olive, & Silk 1990), so an early formation of $\sim 10^{8.5}$ CVs ($\sim 10^{-3}$ of the dark mass) is not an outrageous possibility.

The possible origin of the low-luminosity CV population will be thoroughly investigated elsewhere, but let us briefly mention an interesting scenario that may already be indicated by x-ray observations of globular clusters. Virtually all globulars contain dim sources with typical soft x-ray luminosity of $\sim 10^{31.5-32.5}$ erg/s in the ROSAT band, and the bulk of these sources is fully consistent with being CVs (Grindlay 1994b, and references therein). X-ray and optical observations indicate a presence of $\sim$10-100 CVs per cluster with $L_x \sim 10^{30.5-31.5}$erg/s, and which are optically fainter than disk CVs for a given x-ray luminosity (Grindlay 1994b). There is also an increasing evidence (Grindlay 1994b,c) that globular cluster CVs may have a considerable magnetic fraction, which would indeed be intrinsicly fainter than non-magnetic CVs. The fact that CVs are so over abundant in globulars relative to the disk (by about an order of magnitude) is not surprising since their formation rate is enhanced in dense environments. Di Stefano and Rappaport (1994) studied the number of CVs that should be active in globular clusters at the present epoch as a result of binary formation via tidal capture. In particular, they predicted of order $\sim 100$ CVs per cluster, and that most CVs should have low luminosities of $\sim 10^{31}$ cgs . Thus, observations of globular clusters and theoretical predictions nicely agree with each other, and both imply a relatively large number of CVs with luminosities that coincide with the derived here for the extended Galactic population (Table 1 and §2.3).

The prevalence of low-luminosity CVs in the Galactic halo, though within globular clusters, may suggest that the Galactic population proposed in this paper is due to CVs that were ejected from globular clusters, and so formed an extended Galactic distribution. Although this process may contribute to the proposed Galactic population, it obviously cannot provide the required number of $\sim 10^{8.5}$ CVs. A variation of this idea is that the observed globular clusters are the least fragile (and most lucky) ones in an original distribution of $\sim 10^{6-7}$ clusters, most of which have been disrupted and destroyed, spreading a large number of binaries over an extended Galactic volume. In particular, a thick disk may form from globular clusters on low-inclination orbits which have been disrupted by encounters with giant molecular clouds, as suggested by Grindlay (1984) for the origin of the field LMXBs. This idea has still not been treated in detail in the literature, and it is unclear at this stage whether it is in conflict with any kind of observation.

It is also possible that a substantial fraction of the dark matter in Galactic halos is in the form of dark clusters (*e.g.*, Carr & Lacey 1987; Wasserman & Salpeter 1994; Ben Moore & Silk 1994). Silk (1991) suggested that these clusters may contain a large fraction of neutron stars and thus a high enough number of LMXBs that could be detected in x-ray observations. Modifying Silk's proposal by replacing neutron stars by CVs would suggest that if such stellar dynamical systems do exist (or existed), and they contain(ed) $\gtrsim 10^{-3}$ of their mass in the form of white dwarfs, then they may provide the required number of CVs. The existence of such clusters has not yet been indicated by any observation, nor it can be ruled out at the present. It would, however, be confirmed or ruled out in the near future by the results of gravitational microlensing surveys (Maoz 1994).

### 4.3. A Thick Disk of Neutron Stars?

We wish to point out an alternative origin of the x-ray sources which we find to be a quite unlikely possibility. However, it is an interesting enough idea to be worth being mentioned and be kept in mind.

The neutron star birth-rate at the present epoch is estimated to be $\sim 10^{-1}$-$10^{-2}$ yr$^{-1}$ (*e.g.*, Taylor & Manchester 1977) which, if persisted throughout most of the life of the Galaxy implies a total number of $\sim 10^{8}$-$10^{9}$ neutron stars. Such large Galactic population of old neutron stars *must* exist, and their total number coincides with the predicted total number of $\sim 10^{8.5}$ objects for the proposed population (Table 1). The expected spatial distribution of neutron stars also fits our requirements: there are indications that pulsars may acquire high birth velocities (Lyne, Manchester & Taylor 1985; Lyne & Lorimer 1994; Frail, Goss & Whiteoak 1994) which, when combined with their initial circular velocity in the Galaxy, would result in them being distributed within a thick flattened distribution (*e.g.*, Hartmann, Epstein & Woosley 1990; Paczyński 1990). The main difficulty in associating old neutron stars with the proposed Galactic population is matching the required properties of their x-ray emission ($L_x \sim 10^{30-31}$ erg/s, and a few KeV spectrum).

Neutron stars have been established as the underlying objects of many astrophysical phenomena, but have not yet been seen in the sense of detecting their surface radiation. Helfand, Chanan and Novick (1980) reviewed the prospects of detecting relatively old neutron stars in x-ray observations, and the various heating mechanisms that may maintain surface temperatures of several hundred thousand degrees or higher. These include heating of the polar caps via bombardment of the surface with relativistic particles from the magnetosphere, frictional dissipation from the crust-superfluid interface, micro-starquakes, and accretion from the interstellar medium. Some of the mechanisms are claimed to be capable of producing x-ray luminosities which are consistent with the required $10^{30-31}$ erg/s (see Helfand *et al.* 1980, and references therein), but the predicted spectrum is too soft to be consistent with the inferred for the proposed population (§3). Combined with the fact that the observed pulsars, although much younger than most existing neutron stars, exhibit much softer spectra than required (*e.g.*, Ögelman, Finely, & Zimmerman 1993; Finley 1994), we conclude that it is a major difficulty for the neutron stars hypothesis.

But, it is not impossible that there is a dominant heating mechanism, either one of the above mechanisms or a yet unrecognized one, that could produce the required typical luminosity and spectrum. However, until such mechanism is proposed, the possibility that we may have begun detecting a large number of old neutron stars in deep x-ray observations should be regarded as a very speculative hypothesis.

## 5. CONCLUSIONS AND PREDICTIONS

We have proposed the existence of a large population of intrinsicly faint (and presumably old) cataclysmic variables which are distributed within a thick Galactic disk (or an oblate halo) with a scale height of a few Kpc. The inferred characteristic properties of this population are an x-ray luminosity of $\approx 10^{30-31}$ erg/s, a number density of sources in the Solar vicinity of $\approx 10^{-5}$-$10^{-4}$ pc$^{-3}$, and a total number of $\approx 3\times10^{8}$ sources in the Galaxy. We show that such Galactic population could resolve the discrepancy noticed in the number counts of faint x-ray sources, account for the harder average spectrum of fainter sources, and may explain the inconsistency found in the clustering properties of the unresolved background. At the same time, it is consistent with observational constraints on the total XRB intensity, the number of unidentified bright sources, the total x-ray luminosity of the Galaxy, the results of fluctuation analyses of the unresolved XRB, and with independent observations of nearby CVs. Although the observed XRB anisotropy seems to indicate a flattened distribution, it is impossible to exclude the (less likely) possibility that the proposed population is distributed within a spherically symmetric halo with a core radius of $\gtrsim 10$ Kpc, in which case the local number density of sources would be $\sim 10^{-5.5}$ pc$^{-3}$, and their typical x-ray luminosity would be $\sim 10^{31}$ erg/s. The existence of either type of population would have negligible implications on the Galaxy mass budget, Galactic dynamics, and on gravitational microlensing surveys.

The Galactic origin hypothesis can be tested by searching for a large-scale variation in the surface number density of resolved x-ray sources. We predict that the density of sources with flux below $10^{-14}$ cgs will decrease either with an increasing galactic latitude ($|b|$) or with an increasing angular distance to the Galactic center ($\theta$). The data presented by Hasinger *et al.* (1993) are insufficient for performing this test for two reasons: 1) the analyzed fields differ significantly in exposure time and in the $H_I$ column density along the different lines of sight, and thus do not consist a uniform sample. 2) the number of sources detected in each field is typically between 10 to 40 (see the "clean sample"). Since $\sim$70% of the sources at the average flux limit are AGNs, the expected number of Galactic sources in each field is $\approx$ 3-12, and there would be no escape from dealing with small number statistics. This is an especially difficult problem due to the fact that the directions to these fields do not have an effectively broad distribution either in $b$ or in $\theta$, so the expected number of Galactic sources would not vary significantly from field to field. For example, assuming that a third of the detected sources belong to the proposed Galactic population, we expect an excess of $\simeq$ 10% (6%) in the total number of sources in the HR8905 field $[(l,b)=(99°,-35°)]$ over the QSO1202 field $[(l,b)=(206°,80°)]$ for a thick disk distribution with a scale height of 3 Kpc (6 Kpc). Assuming a spherical halo distribution, we expect an excess of $\simeq$22% (10%) in the $\alpha$ Boo field ($\theta=69.7°$) over the Nower1 field ($\theta=122.1°$) for a core radius of 3 Kpc (15 Kpc).

Therefore, there are three conceivable ways to proceed: 1) establishing a more homogeneous sample of deep and equal exposure-time ROSAT HRI images of a few optimally selected fields, namely, fields which differ significantly in $b$ and $\theta$, but have similar $H_I$ column density (the PSPC is not available any more). In order to maximize the number of fields in such a sample it would be best, if possible, to choose new fields which have $H_I$ column density similar to that of some of the fields already studied by Hasinger *et al.* 2) it might be possible to perform the above test using only the data by Hasinger *et al.* (1993), but it would require taking into account only sources fainter than $10^{-14}$; assuming a logN-logS relation for AGNs and extracting their estimated contribution to the number counts, and normalizing the number of detected sources to equal exposure-time and equal $H_I$ column density. This would require a detailed analysis of the raw data, and it is unclear whether it could provide a conclusive test. 3) an existence of a Galactic population implies also a direction-dependent contribution to the unresolved XRB. It might be possible that fluctuation analyses could detect a large-scale variation in the $P(D)$ distributions of the unresolved background.

We also suggest a test for the proposed nature of the Galactic population, *i.e.*, that these are intrinsicly faint cataclysmic variables. Regardless of whether these are non-magnetic CVs or magnetic ones (*e.g.*, AM-Her CVs), they should appear almost as "pure" emission line sources. The reason for that is that magnetic CVs typically have very faint continuum , and non-magnetic CVs, the fainter they are in the visual band, the more enhanced the lines become relative to the continuum (Patterson 1984; Patterson & Raymond 1985). Therefore, the proposed CV population could, in principle, be detected in an all-sky emission line survey, which may not be a very practical technique. A much more efficient strategy would be to take both deep narrow-band (*e.g.*, H$\alpha$) images as well as broad band (R) images of the EMSS fields (Hasinger *et al.* 1993). Since either type of CVs are expected to be a few times fainter in the continuum than in the H$\alpha$, they should be easily identified in a comparison of the two images. Such test will require a limiting magnitude of $\sim$24 ($M_v\sim+11$ at a distance of a few Kpc).

Our prediction is that $\sim$20% of the x-ray sources with flux $\lesssim 10^{-14}$ cgs in the 0.5-2 KeV band should appear considerably brighter in the H$\alpha$ than in the R band. We emphasize that *all* faint x-ray sources in the EMSS fields should be examined, including ones already "identified" as AGNs since there is a chance that some fraction of these have been misidentified. Afterall, at these flux limits the number density of AGNs should be of order $\approx$200 deg$^{-2}$ which implies an average angular separation of $\sim 4'$. Given the localization accuracy of $\sim 1'$ one cannot exclude the possibility of false identifications.

We hope that these tests could be performed in the near future. Otherwise, a verification of our proposal may have to wait for the AXAF mission. AXAF will have a greater sensitivity, better spectral resolution and better localization which will enable a much more conclusive test for the nature of these sources.

EM wishes to thank Rosanne Di Stefano and Elihu Boldt for the useful discussions. This work was supported by the U.S. National Science Foundation, grant PHY-91-06678, and NASA, grant NAGW-3280.